\documentclass[showpacs,preprintnumbers,amsmath,amssymb,nofootinbib]{revtex4}
\usepackage{color}
\usepackage{graphicx}% Include figure files
\usepackage{dcolumn}% Align table columns on decimal point
\usepackage{bm}% bold math
\usepackage{latexsym}
\usepackage{epsfig}
\usepackage{rotating}

\newcommand{\gdualn}[1]{\overset{\:{}^{{}^{\boldsymbol{\neg}}}}{\smash[t]{#1}}} %Elko dual

\begin{document}

\title{Degeneracy pressure of mass dimension one fermionic fields\\ and the dark matter halo of galaxies}

\author{S. H. Pereira}

\affiliation{Departamento de F\'isica\\Faculdade de Engenharia de Guaratinguet\'a\\ Universidade Estadual Paulista (UNESP) \\  Av. Dr. Ariberto Pereira da Cunha 333\\
12516-410 -- Guaratinguet\'a, SP, Brazil}

\begin{abstract}
In this paper, the recently proposed mass dimension one fermionic field is supposed to be responsible for the dark matter halo around galactic nuclei, through the quantum degeneracy pressure effect of the field. It will be showed that the mass-ratio relation for dwarf galaxies can be well explained for a particle dark matter mass of about $100$eV - $200$eV. For a large galaxy, as Milky Way, the observational data for rotation curve can be well reproduced for a particle mass of about 23eV, with the addition of other substructures.

\end{abstract}

\maketitle

\section{Introduction}

Since the 2005 formulation of a new class of fermionic fields with mass dimension one~\cite{Ahluwalia:2004sz,Ahluwalia:2004ab}, initially named Elko, and its recent establishment on solid grounds \cite{Ahluwalia:2016rwl,Ahluwalia:2019etz}, several works have studied its consequences and applications on different approaches, including cosmological scenarios~\cite{Pereira:2020ogo,Pereira:2018hir,Pereira:2017efk,Pereira:2016emd,Pereira:2014wta,daSilva:2014kfa,Boehmer:2007dh,Boehmer:2010ma,Boehmer:2008ah,
Boehmer:2009aw,Boehmer:2007ut,Boehmer:2008rz,Villalobos:2018shc,Sadjadi:2011uu,Basak:2011wp,Lee:2012qq,Basak:2012sn,Basak:2014qea}, Hawking radiation and black holes ~\cite{daRocha:2014dla,
 Cavalcanti:2015nna,daSilva:2016htz}, Casimir effect, particle creation and thermodynamic properties~\cite{Maluf:2019ujv,Pereira:2016muy,Pereira:2016eez,Pereira:2018xyl}, among others~\cite{Fabbri:2017xyk,Fabbri:2014foa,Bernardini:2012sc}. The two main features of this new kind of field are that they are neutral spin-$1/2$ fields constructed from charge conjugation spinors and have {canonical mass dimension one which prohibits them} to couple to other particles of the {Standard Model} at tree level\footnote{Although not interacting with {Standard Model} particles, the interaction of the mass dimension one field with the Higgs field at tree
level through a fourth interaction is possible. The interaction with photons (gauge fields) and other {Standard Model} spinors occurs only at 1-loop order \cite{Alves:2017joy}.}. These features make such field a natural candidate to describe dark matter (DM) particles in the universe. 

The first and main evidence for the presence of dark matter around galactic nuclei come from the observation of galactic rotation curves. From the Kepler Second Law, it is expected that the rotation velocities of stars will decrease with distance from the centre,  however the observations show that most galaxy rotation curves remains flat as distance from the centre increases. This evince the presence of an invisible {DM halo} component around galactic nuclei.

Although the {Standard Model} of cosmology, namely $\Lambda$CDM model, predict about $25\%-30\%$ of cold dark matter (characterised by particles of mass $\sim$ 100GeV ) in the composition of the universe at large scale, {local
observations of dark matter on galactic scales point towards the necessity of lighter particles ($\sim$ keV), called warm dark matter, to correctly explain} the density profile and rotation curves of galaxies. Cold dark matter models produces an overabundance of substructures below scales of about 50 kpc while warm dark matter models produces the desired amount of structures at these scales. Additionally, cold dark matter models predicts a steep density profile in dwarf galaxies, contrary to smooth DM profiles at the
galactic cores, which can be realised only for sufficiently light DM particles. In order to solve these problems, models of fermionic warm dark matter have been studied recently in the context of a quantum system.

It is well known that at low temperature a degenerate gas of free fermions exhibits a quantum pressure due to Pauli exclusion principle. Such pressure have been advocated recently as responsible for dark matter halo of galaxies, where the gravitational collapse of the gas is prevented due to degeneracy pressure \cite{Destri:2013pt, deVega:2013jfy, Domcke:2014kla, Randall:2016bqw, Pal:2019tqq}. Particularly, dwarf spheroidal galaxies are supposedly dominated by DM particles, which makes these objects good laboratories for testing such model. Larger spheroidal and elliptic  galaxies can also be considered with some additional suppositions concerning its substructures.

Domcke \& Urbano \cite{Domcke:2014kla} have used a completely degenerate model to constrain the DM mass of dwarf galaxies in the range 100eV $\sim$ 200eV, and Randall et al. \cite{Randall:2016bqw}  treat the
dark matter as a quasi-degenerate Fermi gas surrounded by a thermal envelope, which reduce these constraints to $m_f \gtrsim 50$eV,
with no upper bound on the mass. An important limit of $m_f\gtrsim 530$eV was set by Baur et al. \cite{Baur:2015jsy}, probed by the Lyman-$\alpha$ transition of distant quasars. However, such large bound can be relaxed at the cost of more complicated dark matter model, as a flavoured
dark matter model or if the dark matter has a non-thermal momentum distribution \cite{Randall:2016bqw}. A conservative lower bound on the mass of fermionic dark matter of 70eV was found at \cite{Randall:2016bqw}. A quite more elaborated model was studied by Destri et al. \cite{Destri:2013pt}. Using a slightly different approach and data from a set of large
elliptical and spiral galaxies and from a small set of dwarf galaxies, Pal et al. \cite{Pal:2019tqq} found that their model can
explain most of the bulk galactic properties, as well as some of the features observed in the rotation
curves, provided the DM mass is in the range of $\sim$ 50eV. J. Barranco et al. \cite{Barranco:2018gjg} found  $29$eV $<m_f<33$eV by constraining light fermionic dark matter with Milky Way observations.

In what follows, after a brief review on the main results for obtaining the thermodynamic properties of the {mass dimension one} fermionic field, we will adopt the mass dimension one fermionic particle as the candidate to DM particle, assuming it is a  non-interacting light free fermion which forms a dark matter halo around galaxies due to its quantum degeneracy pressure.

\section{Partition function of mass dimension one fermionic fields}

The study of thermodynamic properties of mass dimension one  fermionic fields was first done by \cite{Pereira:2018xyl} at null chemical potential, by using techniques from finite temperature field theory \cite{Bellac:2011kqa,Kapusta:2006pm,Das:1997gg,Bailin:1986wt}. The equilibrium thermodynamic properties of systems formed by scalar, fermionic and gauge fields can be extracted from the partition function $Z$ of the system, obtained through a functional integration method and the imaginary time formalism. The equilibrium partition function $Z$ for a fermionic field $\psi$ is done by introducing the temperature by means of the imaginary time formalism, making a rotation from real time axis to complex one, and introducing the new variable $\tau = it$. For a system at equilibrium, the thermodynamic temperature $\bar{T}$ is introduced by means of $\tau \equiv \beta = \frac{1}{k_B \bar{T}}$ and time integration must be taken over the interval $0<\tau <\beta$.\footnote{Just for future reference we are representing the temperature of the field as $\bar{T}$ to distinguish it from the temperature $T$ of {Standard Model} particles in equilibrium in a thermal bath for instance, as cosmic radiation background temperature. Once the mass dimension one fermionic fields do not interact with particles of {Standard Model}, may be such equilibrium is not allowed. This is important for cosmological applications discussed at the end.}

The partition function for a generic fermionic field $\psi(x)$ can be written as \cite{Kapusta:2006pm,Bailin:1986wt}:
\begin{equation}
Z=\int[D\mathfrak{\pi}]\int_{a.p.}  [D\mathfrak{\psi}]\exp\Bigg[\int_0^\beta d\tau \int d^3x\,\bigg(i\pi\frac{\partial\mathfrak{\psi} }{\partial\tau}-\mathfrak{H}(\mathfrak{\psi},\mathfrak{\pi}) + \mu \mathfrak{N}(\mathfrak{\psi},\mathfrak{\pi}) \bigg)\Bigg]\,,\label{eq:pe01}
\end{equation}
where $\mathfrak{H}(\mathfrak{\psi},\mathfrak{\pi}) = \pi \dot{\mathfrak{\psi}}-\mathfrak{L}(x)$ is the Hamiltonian density of the fermionic field $\mathfrak{\psi}$ and its momentum conjugate $\mathfrak{\pi}$,  $\mathfrak{N}$ stands for a conserved ``charge" density to which it is associated a chemical potential $\mu$. The functional integral over $\psi$ must be done with anti-periodic (a.p.) boundary condition:
\begin{equation}
\mathfrak{\psi}(\tau=0,{\vec{x}})=-\mathfrak{\psi}(\tau=\beta,{\vec{x}})\,.\label{eq:pe02}
\end{equation}

Now, let us consider the free field Lagrangian density for a mass dimension one fermionic field ${\mathfrak{f}}(x)$:
\begin{equation}
\mathfrak{L}(x) =\frac{1}{2}\partial^\mu\gdualn{\mathfrak{f}}(x)\,\partial_\mu {\mathfrak{f}}(x) - \frac{m_f^2}{2} \gdualn{\mathfrak{f}}(x) \mathfrak{f}(x), \label{eq:pe03}
\end{equation}
where $\gdualn{\mathfrak{f}}(x)$ is the dual of the field {and $m_f$ is the mass of the fermionic particle}. The momentum conjugate to the field and its dual are $\mathfrak{\pi}(x)=\frac{1}{2}\dot{\gdualn{\mathfrak{f}}}(x)=\frac{i}{2}\partial_\tau\gdualn{\mathfrak{f}}(x)$ and $\gdualn{\mathfrak{\pi}}(x)=\frac{1}{2}\dot{\mathfrak{f}}(x)=\frac{i}{2}\partial_\tau\mathfrak{f}(x)$, respectively. {Although the mass dimension one fermionic field represents a neutral particle, its Lagrangian density admits a U(1) conserved charge $\mathfrak{N}=i \Big(\gdualn{\mathfrak{f}} \gdualn{\mathfrak{\pi}} - \mathfrak{\pi} \,\mathfrak{f} \Big)$, which we interpret as a ``dark" charge  associated to the field , other than electric charge. Thus, just for maintain the generality and parallel to standard Dirac fermions, we introduce a chemical potential $\mu$, ensuring its number conservation.\footnote{{At the end of calculations the chemical potential can be set to zero. The partition function for the specific case of null chemical potential was calculated in  \cite{Pereira:2018xyl}.}}} The important result is: 
\begin{eqnarray}
\ln Z&=&2V\int\frac{d^3p}{(2\pi)^3}\Bigg[\beta \sqrt{ p^2+m_f^2}+ \ln\bigg(1+\exp(-\beta \sqrt{{p}^2+m_f^2}+\mu)\bigg)\nonumber\\
&& + \ln\bigg(1+\exp(-\beta \sqrt{{p}^2+m_f^2}-\mu)\bigg)\Bigg]\,,\label{eq:pe12}
\end{eqnarray}
{where the integral is over the 3-momentum, with $p = |\vec{p}|$.}

The Helmholtz free energy is:
\begin{equation}
F=-\frac{1}{\beta}\ln Z \,,\label{eq:pe13}
\end{equation}
and the main thermodynamic quantities, as total energy $E$, pressure $P$ and entropy $S$, are obtained from the Helmholtz free energy:
\begin{equation}
E=F+\bar{T}S,\hspace{0.5cm} P=-\Bigg(\frac{\partial F}{\partial V}\Bigg)_{\bar{T}},\hspace{0.5cm} S=-\Bigg(\frac{\partial F}{\partial \bar{T}}\Bigg)_V\,.\label{eq:pe14}
\end{equation}

Taking just the temperature dependent part of (\ref{eq:pe13}) and performing an integration by parts we obtain:
\begin{equation}
\frac{F}{V}=-\frac{1}{3\pi^2}\int_0^\infty\frac{p^4}{\sqrt{p^2+m_f^2}}\bigg[\frac{1}{{e}^{\beta(\sqrt{p^2+m_f^2}-\mu)}+1} + \frac{1}{{e}^{\beta(\sqrt{p^2+m_f^2}+\mu)}+1}\bigg]dp\,,\label{eq:pe15}
\end{equation}
where $V=\int d^3x$. The Fermi-Dirac distribution function, $n_\varepsilon\equiv [{e}^{\beta(\varepsilon-\mu)}+1]^{-1}$, with relativistic energy $\varepsilon=\sqrt{p^2+m^2}$ and chemical potential $\mu$ is evident.

%\section{High temperature limit and relativistic degrees of freedom in early universe}

The high temperature limit $(\bar{T}>>m)$ of (\ref{eq:pe15}) can be easily calculated \cite{Bellac:2011kqa,Kapusta:2006pm,Das:1997gg,Bailin:1986wt} and the results for the Helmholtz free energy density, energy density, pressure and entropy density from (\ref{eq:pe14}) are:
\begin{equation}
\frac{F}{V}=-\frac{7\pi^2\bar{T}^4}{180}, \hspace{0.5cm}\frac{E}{V}=\frac{7\pi^2\bar{T}^4}{60}, \hspace{0.5cm} P =\frac{7\pi^2\bar{T}^4}{180}, \hspace{0.5cm}\frac{S}{V}=\frac{14\pi^2\bar{T}^3}{90}.
\end{equation}
%DVA: that -> the
Surprisingly, these results are exactly the same as obtained for a Dirac fermionic field\footnote{The above treatment and main results for thermodynamic properties and high temperature limit were first obtained in \cite{Pereira:2018xyl}.}.

\section{Low temperature limit, degeneracy pressure and dark matter halo of galaxies}

The low temperature limit is interesting to study from an astrophysical perspective, and will be explored here in the context of dark matter in galactic halos. As mentioned above, some works consider the dark matter particles as generic fermionic fields satisfying just some few properties, with no reference to a specific candidate, {such as sterile neutrinos}, gravitinos or weakly interacting massive particles (WIMPs). This kind of light DM candidates {cannot be} in thermal equilibrium with other particles during the large scale structure formation epochs and
big-bang nucleosynthesis. Such property can be achieved if the
DM fermions are endorsed with a conserved charge so that there
are no renormalisable couplings between the DM fermions and the {Standard Model} particles  \cite{Pal:2019tqq}. As pointed out by Randall et al. \cite{Randall:2016bqw}, the possible problem concerning the relic abundance can be alleviated when the dark matter is not in thermal contact with baryons, being colder when
dark matter is decoupled and not reheated by the {Standard Model} particles, for example. These are exactly the main characteristics satisfied by the mass dimension one fermionic field treated here.

Here we will follow the most simplified approach of a completely degenerate gas of reference \cite{Domcke:2014kla} to briefly present the main features of the model.

The equation for the hydrostatic equilibrium between gradient of pressure $P$ and gravitational attractive force is given by
\begin{equation}
    \frac{1}{r^2} \frac{d}{dr} \left(\frac{r^2}{\rho} \frac{dP}{dr} \right) = -4\pi G \rho\,,\label{eq:pe21}
\end{equation}
where $\rho(r)$ is the mass density {distribution, or density profile}. Given a relation $P = P(\rho)$, the above equation can be solved for the mass density $\rho(r)$. In general we are interested in the following conditions
\begin{equation}
    \rho(r = 0) = \rho_0, \hspace{1cm}   \left.\frac{d\rho}{dr}\right|_{r = 0} = 0.\label{cc}
\end{equation}

Additionally, the total mass $M$ of the system is related to its characteristic radius $R$ by
\begin{equation}
M(R)=4\pi\int_0^R r^2 \rho(r) dr.\label{eq:pe23}
\end{equation}

\begin{figure}[t]
\centering
\includegraphics[width=0.50\textwidth]{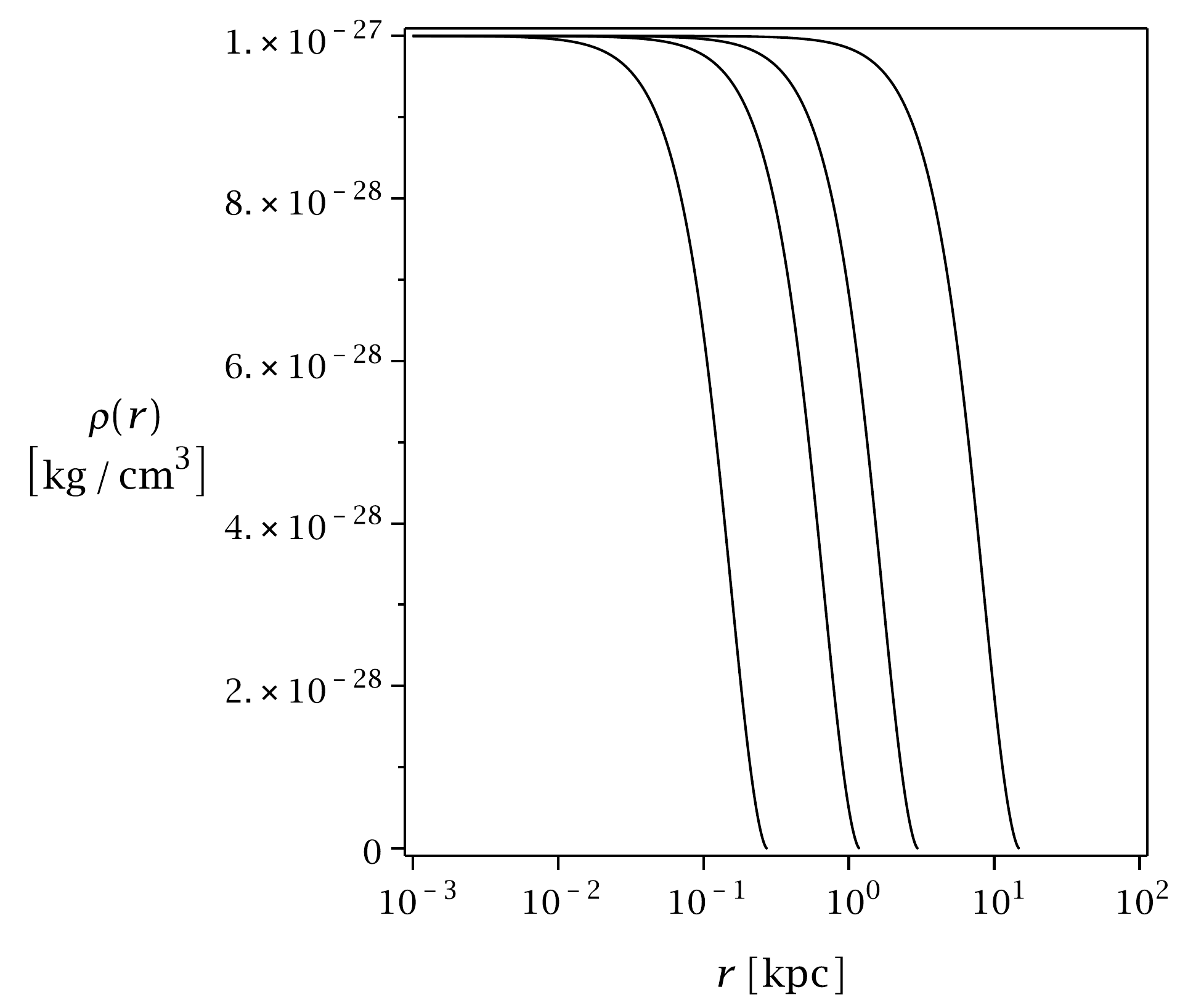}
\includegraphics[width=0.46\textwidth]{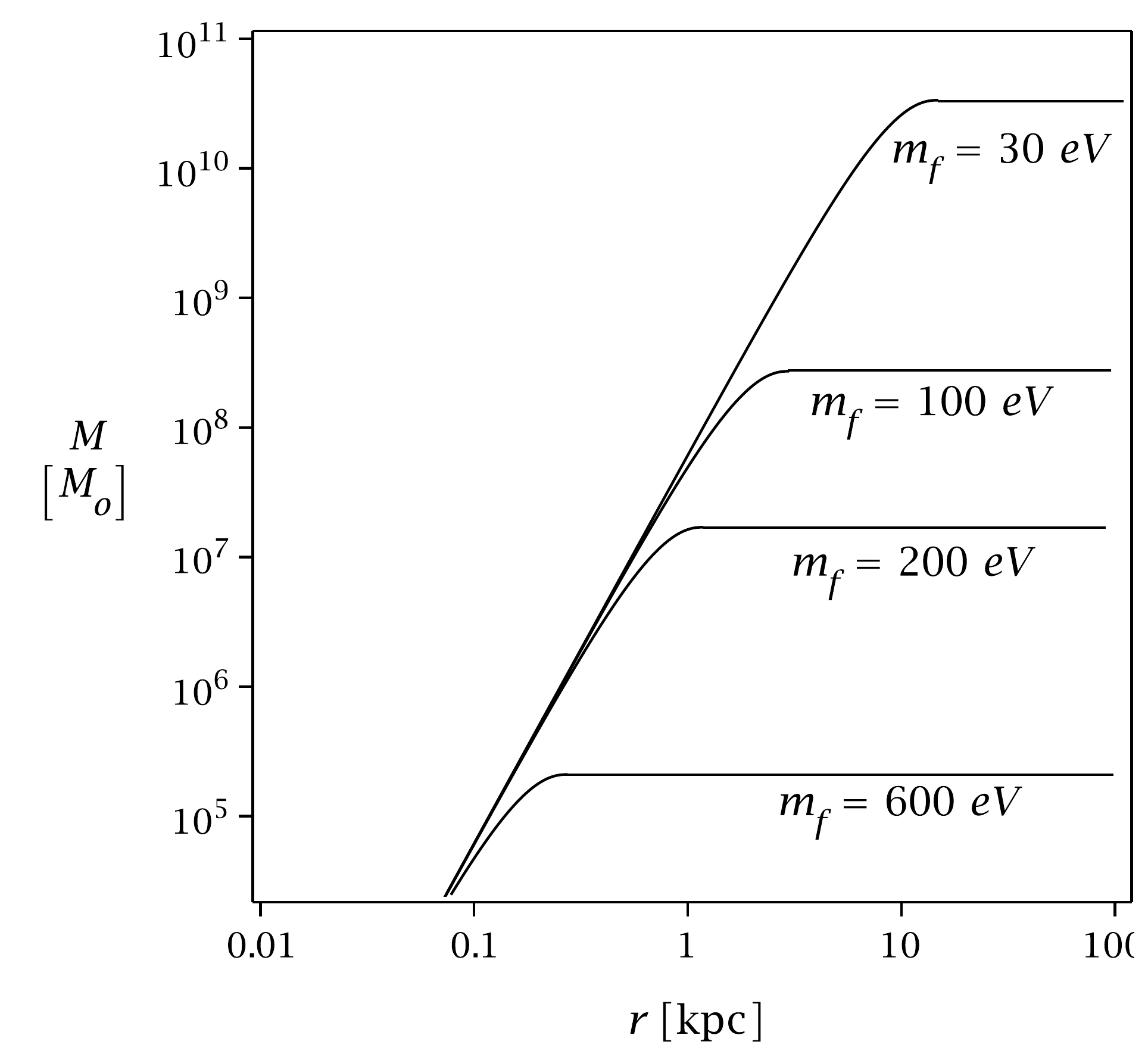}
\caption{(Left) Density profile for different values of the particle mass $m_f$. From left to right: 600eV, 200eV, 100eV and 30eV, for a central density $\rho_0 = 1.0\times 10^{-27}$kg/cm$^3$. (Right) Total mass for different values of the particle mass, for a central density of $10^{-27}$kg/cm$^3$. }\label{fig023pe}
\end{figure}

{Next we will solve (\ref{eq:pe21}) subject to (\ref{cc}) for a degenerate gas subjected just to a degeneracy pressure $P_d$}, corresponding to the low temperature limit, $\bar{T}\sim 0$. The Fermi-Dirac distribution function (\ref{eq:pe15}) turns into a step function equal 1 for $p\leq p_F$ and 0 for $p > p_F$, where $p_F = (3h^3\rho/8\pi m_f)^{1/3}$ is the Fermi momentum associated to a particle of mass $m_f$. The pressure integral coming from (\ref{eq:pe14})-(\ref{eq:pe15}) is\footnote{We have reintroduced the Planck constant and light velocity just for dimensional analysis.}:
\begin{equation}
P_d=\frac{1}{3\pi^2}\bigg(\frac{2\pi}{h}\bigg)^3\int_0^{p_F}\frac{p^4}{\sqrt{p^2/c^2+m_f^2}}dp\,.\label{eq:pe24}
\end{equation}
For a non-relativistic particle, $p_F \ll m_fc$, the integral gives the well known result:
\begin{equation}
P_d=\frac{h^2}{5m_f^{8/3}}\bigg(\frac{3}{8\pi}\bigg)^{2/3}\rho^{5/3}.\label{eq:pe25}
\end{equation}
This is the equation of state relating the degeneracy pressure to mass density $\rho$, the starting point to solve (\ref{eq:pe21}), given a fermionic mass $m_f$ and a central density $\rho_0$. Figure \ref{fig023pe} shows the density profile (Left) and total mass (Right) distributions as a function of the radius for a degenerate Fermi gas described
by the equations (\ref{eq:pe21})-(\ref{eq:pe23}), for a typical central density $\rho_0 = 1.0\times 10^{-27}$kg/cm$^3$.
Observe that the density profile is nearly constant up to a characteristic radius and decreases abruptly to zero. Also, larger the value of mass lesser the limit radius. The total mass increases abruptly and is constant after the limit radius, varying from about $10^5 M_\odot$ for $m_f = 600$ eV up to about $5\times 10^{10} M_\odot$ for $m_f = 30$ eV. The range from $10^6 M_\odot - 10^8 M_\odot$ is typical for some dwarf galaxies, while the range $> 10^{10} M_\odot$ is typical for middle and large galaxies, as we shall see in next analysis.

With a change of variable $\rho=\rho_0\theta^{3/2}(\xi)$ and re-scaling the radial coordinate as $\xi=r/\alpha$, where $\alpha = \sqrt{5K/8\pi G \rho_0^{1/3}}$ and $K=(h^2/5m_f^{8/3})(3/8\pi)^{2/3}$, equation (\ref{eq:pe21}) can be written as the well known Lane-Emdem equation,
\begin{equation}
    \frac{1}{\xi^2}\frac{d}{d\xi}\bigg(\xi^2\frac{d\theta}{d\xi}\bigg)=-\theta^{3/2}\,,\label{eq:pe26}
\end{equation}
 which can be solved numerically using the boundary conditions $\theta(0) = 1$, $\theta'(0)=0$. The first zero of the solution for $\theta(\xi)$ occurs at $\xi_1 \simeq 3.65$, which indicate a maximum radius ${R}=\xi_1 \alpha$ above which the matter density is null and the total mass inside the volume is given by (\ref{eq:pe23}). These quantities are strongly dependent on the central density $\rho_0$ and fermion mass $m_f$, and they can be put into a mass-radius relation:
 \begin{equation}
     M=4\pi \xi_1^5 | \theta'(\xi_1)|\bigg(\frac{5K}{8\pi G}\bigg)^3\frac{1}{R^3}\,,\label{eq:pe27}
 \end{equation}
 with $\theta'(\xi_1)\simeq -0.2037$ from numerical estimates. We see that, {since $K \sim m_f^{-8/3}$}, for a fixed fermionic mass $m_f$, greater the radius smaller the total mass inside $R$. On the other hand, for a fixed radius $R$, smaller the fermionic mass greater the total mass $M$.
 \begin{figure}[t]
\centering
\includegraphics[width=0.75\textwidth]{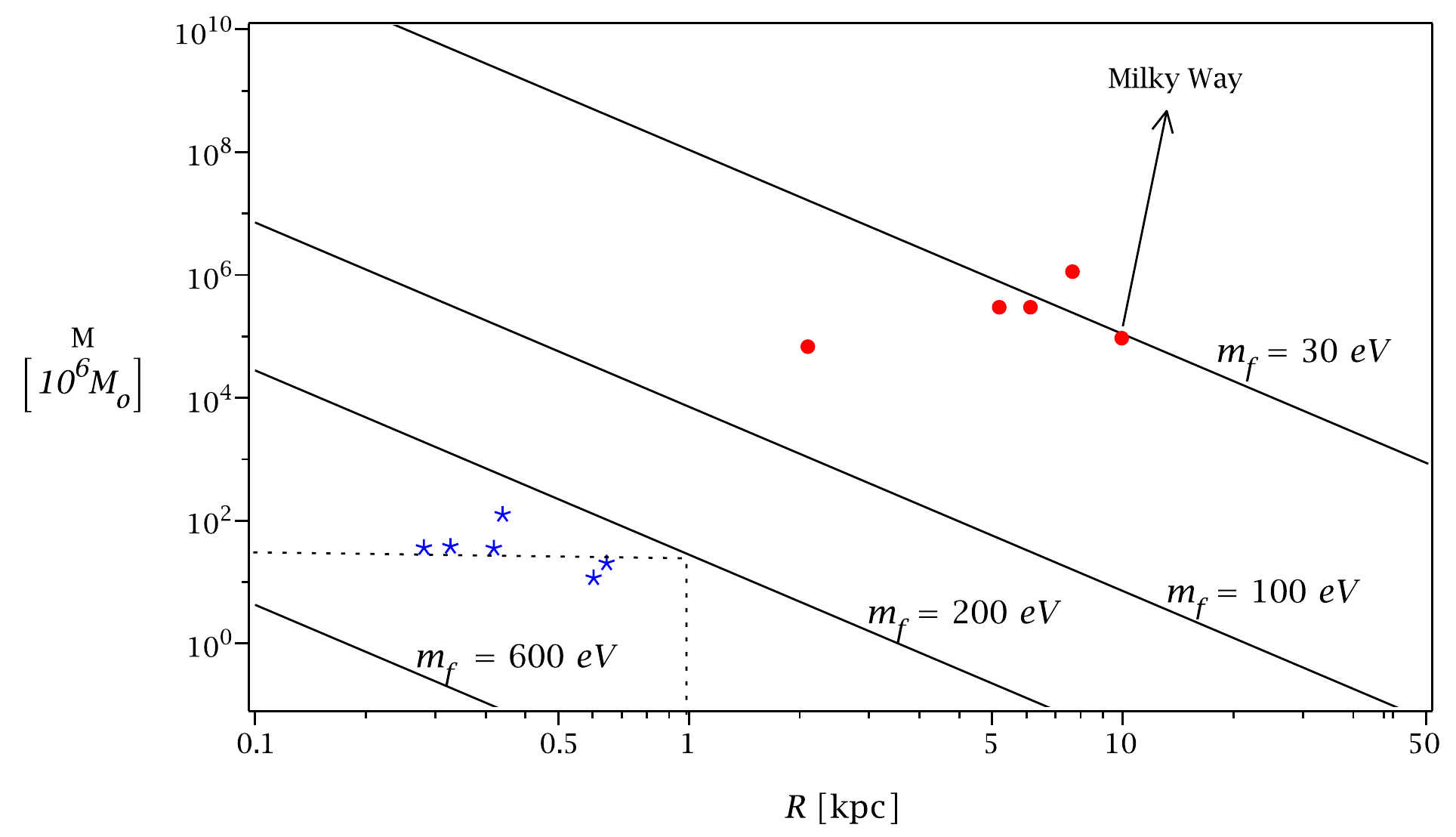}
\caption{Mass-radius relation (\ref{eq:pe27}) for $m_f = 30$eV, $m_f = 100$eV, $m_f = 200$eV and $m_f=600$eV and observational data for six classical dwarf spheroidal galaxies (blue stars) and five middle to large spiral galaxies (red circles), from left to right: Draco, Carina, Sculptor, Leo I, Ursa Minor, Fornax, NGC 4478, NGC 3853, NGC 731, NGC 499 and Milky Way. Data are taken from
refs. \cite{deVega:2013jfy} and \cite{Sofue:2016kja}. }\label{fig01pe}
\end{figure}

Figure \ref{fig01pe} shows the mass-radius relation (\ref{eq:pe27}) ({straight lines}) for different values of fermion mass, namely $m_f = 30$eV, $m_f = 100$eV, $m_f = 200$eV and $m_f=600$eV, {where $M$ is the total dark matter mass of galaxy and $R$ corresponds to its dark matter halo.} In blue stars we present the observational data for total mass and radius for six classical dwarf spheroidal galaxies, with cores from 245pc to 646pc and masses $10^7 M_\odot - 10^8 M_\odot$. Taking for instance an average mass of about $5\times 10^7 M_\odot$ (horizontal doted line) we obtain a radius of $1$kpc (vertical doted line) to the dark matter halo of these objects if $m_f=200$eV. It is greater than the observed radius for these type of galaxies.  It is quite interesting that this very simplified model can satisfactorily explain the presence of a DM halo for such kind of galaxies exclusively due to degeneracy pressure of mass dimension one fermionic particles, with a mass of few hundred eV. A complete study of the projected velocity dispersion for the above cited six classical dwarf spheroidal galaxies and others are given in references \cite{Domcke:2014kla,Randall:2016bqw}, which point for a mass of about 100eV - 200eV.

%DVA: Milk->Milky
Figure \ref{fig01pe} also shows the observational data for {total} mass and {the corresponding dark matter radius} of five middle to large spiral galaxies (red circles). Contrary to dwarf galaxies the dark halo of these objects {requires} a much smaller particle mass in order to explain their
large halos maintained just by degeneracy pressure. Taking for instance the Milky Way, a galaxy of about $10^{11}M_\odot$, a fermionic particle mass of about $30$eV guarantees a dark matter halo up to about $10$kpc, which is lesser than its visible radius, of about 27kpc. The total dark matter halo radius is estimated to extend far beyond that. 

The main observational evidence for the need of a dark matter halo around the galactic nuclei comes from the rotation curves of galaxies. Contrary to dwarf galaxies, which do not rotate, it is well known that presence of dark matter in large galaxies drastically alters the dynamic of rotation of stars around its centre, mainly at large distances. However, due to much more complex structure of large galaxies, rotation curves cannot be explained just by the dark matter component.

%DVA: Milk->Milky
It is well known that large galaxies have a much more complex structure, characterised by different {mass scales} $M_i$ and radius $r_i$. Taking for instance the Milky Way, it has five main structures:

\emph{1 - Black Hole}: a central black hole of $M_{BH} \simeq 3.6\times 10^6 M_\odot$, that strongly influences the rotation velocity of stars for $r \lesssim 1$pc;

\emph{2 - Massive Core}: a first massive core bulge with $M_C \simeq 4.0\times 10^7 M_\odot$ and $r_C \simeq 3.5$pc, which acts on scales up to order of $\sim$ 10pc;

\emph{3 - Main Bulge}: a second massive bulge with $M_B \simeq 9.2\times 10^9 M_\odot$ and $r_B\simeq 120$pc, which acts on scales up to few hundred pc;

\emph{4 - Disk}: a large disk up to few kpc and a mean mass of about $10^{10}M_\odot$. The disk can also contain arms, rings, bar and interstellar gas, thus its dimensions vary depending on the specific model adopted;

\emph{5 - Dark Matter}: a dark matter halo of about $M_{DM} \simeq 1.0\times 10^{11} M_\odot$ and $r_{DM} \simeq 10$kpc, which actually continues for several dozen kpc and is not visible.
%DVA: taking -> taken, taking-> taken

The above values were taken from \cite{Sofue:2016kja}. All these components must be taken into account when calculating the rotation curve of a galaxy. The rotation velocity of a star at a distance $r$ of the galactic centre is given by:
\begin{equation}
V_i(r) = \sqrt{\frac{GM_i(r)}{r}}\,,\label{eq:pe28}
\end{equation}
where $M_i(r)$ is the total mass of the component $i$ at position $r$. By considering just the dark matter component formed by mass dimension one fermionic particles with an estimate mass $m_f = 23$eV and a central density of $3.0\times 10^{-27}$g/cm$^3$, the set of equations (\ref{eq:pe21}) - (\ref{eq:pe23}) can be solved to obtain $M_{DM}(r)$. A total mass of $M_{DM} = 1.7\times 10^{11} M_\odot$ is reached at $r_{DM} \simeq 16$kps, about twice the distance from centre to the Sun. Such total mass value is compatible with the estimated total mass for Milky Way.

\begin{figure}[t]
\centering
\includegraphics[width=1.0\textwidth]{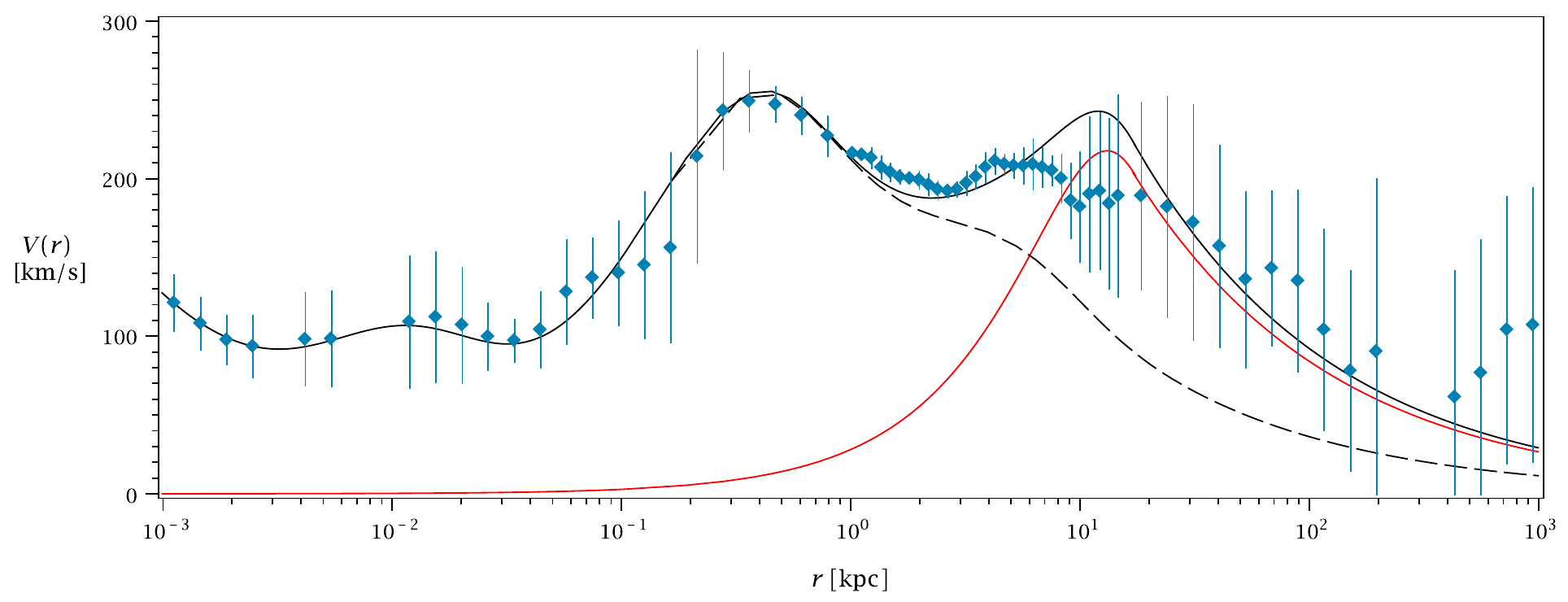}
\caption{Rotation curve for Milky Way (blue points, with error bars taken from \cite{Sofue:2013kja}). Theoretical velocity obtained just by degenerate gas of mass dimension one fermionic particles (red line) with $m_f=23$eV and theoretical velocity (black line) obtained with the inclusion of a central black hole, two core bulges, a thin disk and mass dimension one particles. In black dashed line the model without mass dimension one particles, for comparison.}\label{fig04pe}
\end{figure}

%DVA: Milk->Milky ABOVE/BELOW
Figure \ref{fig04pe} shows the observational data (blue points with error bars) for rotation curve of Milky Way from $1$pc to $10^6$pc, obtained from \cite{Sofue:2013kja}. In solid red line we plot the theoretical value obtained for the rotation velocity caused just by the dark matter halo of degeneracy pressure of fermionic particles with $m_f=23$eV and a central density of $3.0\times 10^{-27}$g/cm$^3$. It is important to emphasise that the shape of the curve is very sensitive to small changes in mass and central density. Although this very simplified model correctly reproduces the observed velocity of about $240$km/s at a distance of about $10$kpc, it is evident that the model fails to describe the regions $r \lesssim 10$kpc and $r \gtrsim $ 1000kpc.

However, as stated before, large galaxies have additional structures that must be considered in the calculation of its rotation curve. In this model we will include all the five structures listed above for Milky Way. The first structure is a black hole. The second and third are the massive core and main bulge of baryonic and luminous matter. Different phenomenological baryon density profiles have been used in order to describe the mass distribution of spiral and elliptical galaxies, as Plummer, Hernquist and Jaffe profiles \cite{Pal:2019tqq}, or Exponential Spheroidal, Semi-isothermal spherical distribution and Vaucouleurs profiles \cite{Sofue:2016kja}, among others. Here we will use just a simple exponential model of the form $\rho_i(r) = \rho_{i0}\exp(-r/r_i)$, where $r_i$  is a characteristic radius and $\rho_{i0}$ represents its central density. We will use the values $r_C$ and $r_B$ from the above list, where the total masses $M_C$ and $M_B$ are obtained by $M_i=4\pi\int_0^\infty r^2 \rho_i(r) dr$. For the disk we use directly the rotation curve of a thin exponential disk \cite{Sofue:2013kja}, with total mass $M_D = 2\pi \rho_{0D}r_D^2$ and rotation velocity expressed by
\begin{equation}
V_D(r) = \sqrt{4\pi G \rho_{0D} r_D y^2[I_0(y)K_0(y)-I_1(y) K_1(y)]}\,,\label{eq:pe29}
\end{equation}
where  $y = r/(2r_D)$, and $I_j$ and $K_j$ are the modified Bessel functions. We use the values $r_D = 2.0$kpc and $M_D = 7.8\times 10^{9}M_\odot$.  It is important to stress that the rotation curve for the disk is affected by additional masses due to arms, as is the case for
 Milky Way, rings, bars and
interstellar gas, which are not being considered in this simplified model.

The total rotation velocity can be written as a superposition of each mass components:
\begin{equation}
   V(r) =  \sqrt{\sum_i V_i^2(r)}\,,\label{eq:pe30}
\end{equation}
with $V_i(r)$ given by (\ref{eq:pe28}) for $i=BH,\, C,\,B,\, DM$ and $V_D$ from  (\ref{eq:pe29}).
The black line in Figure \ref{fig04pe} shows the effects of the five structures on the rotation velocity for Milky Way. It is quite evident that this model correctly reproduces the observational data in the whole region from 1pc to $10^6$pc. It is also evident that the main contribution from the degeneracy pressure of the fermion field is for $r\gtrsim 2$kpc. Just for comparison, we have showed in black dashed line the rotation curve without the dark matter component. The region $1 - 10$kpc is strongly affected by the specific disk model adopted. For the region $r\gtrsim 10$kpc the main contribution comes from dark matter, and continue for several hundred pc.

The flatness of the overall shape of entire rotation curves is a general feature of spiral and elliptic galaxies. In the specific case of Milky Way, the region $r \gtrsim 400$kpc also present such behaviour and the present model fail to explain the observational data. The main reason for the decrease of the velocity rotation for $r \gtrsim 20$kpc is the abrupt decreasing of the density profile due to properties of the completely degenerate gas, which corresponds to the limit $T \to 0$, when the Fermi-Dirac distribution function turns into a step function. A small deviation from the degenerate case changes the equations (\ref{eq:pe24}) and (\ref{eq:pe25}), once that the total pressure must be calculated with the complete expression (\ref{eq:pe15}), and the momentum integral goes up to infinity. This also introduces a new free parameter, namely the temperature $T$ of the gas, which also increases the energy of the whole system and breaks the condition of an abrupt cut off for the density profile, allowing for a slight growth of the rotation velocity in the region $r \gtrsim 20$kpc.

A complete study of this problem
is of course beyond the purposes of this review,  nevertheless it is interesting to provide at least some quantitative aspects. For a gas of fermions,
the degeneracy temperature $T_d$ that defines whether the system can be treated as degenerate or not can be written as \cite{Domcke:2014kla}:
\begin{equation}
    T_d = \frac{h^2}{2\pi m_f k_B}\bigg(\frac{\rho_0}{2m_f}\bigg)^{2/3}\,,
\end{equation}
which furnish $T_d \simeq 0.130$K for the values $m_f = 23$eV and $\rho_0 = 3\times 10^{-27}$kg/cm$^3$ used in the analysis. It is well known that the limit $T\gg T_d$ corresponds to a Maxwell-Boltzmann distribution function, so that even for a temperature of the order of the cosmic microwave background (CMB) temperature of about $2.7$K this model must be better studied in view of a non-degenerate fermion gas. However it is well known that mass dimension one fermionic particles do not interact with electromagnetic field {at tree level}, thus its temperature evolution can be {very different} from CMB.

The addition of these new ingredients makes the problem quite complicated and beyond the scope of this review. What we have done here is show that for both dwarf and large galaxies the degeneracy pressure of mass dimension one fermionic particles may be responsible for the existence of a {DM halo} around these galaxies. {For dwarf galaxies, some recent studies} for general fermionic particles {point to a mass range} of 100eV $\sim 200$eV , and even lower masses are possible for models such that dark matter constitutes a hidden sector, essentially decoupled thermodynamically from % DVA Milk->Milky
the visible sector \cite{Randall:2016bqw}. This is the case for mass dimension one fermionic particles. For large galaxies, specifically the Milky Way, the degeneracy pressure makes a significant contribution to the rotation curve of the galaxy for a mass $\lesssim$ 23eV, with the inclusion of other substructures. Such mass value is consistent with the one obtained in \cite{Barranco:2018gjg}.

\section{Conclusion}

The thermodynamic properties of  the recently proposed mass dimension one fermionic field and the supposition that it must be responsible for the dark matter halo around galactic nuclei were studied. Through the quantum degeneracy pressure effect at low temperature, {is was shown} that the mass-ratio relation for dwarf galaxies can be well explained for a particle dark matter mass of about 100eV - 200eV. {For a large galaxy like the Milky
Way, by adding substructures such as a central black hole, main core, bulk and disk, its
rotation curve can be reproduced for a particle mass of about 23eV}. The main contribution from the dark matter halo for the rotation curve is for a radius from 1 to $10^3$kpc. The inclusion of non-degenerate effects must increase this range. Also, the inclusion of a self-interaction term to the Lagrangian density (\ref{eq:pe03}) also must give additional contributions to the above discussions. However, the functional integration in obtaining the partition function for the field is not analytic with a self-interaction term and some other method must be adopted to obtain the thermodynamic properties of the field.

\begin{acknowledgements}
This study was financed in part by the Coordena\c{c}\~ao de Aperfei\c{c}oamento de Pessoal de N\'ivel Superior - Brasil (CAPES) - Finance Code 001. SHP would like to thank CNPq - Conselho Nacional de Desenvolvimento Cient\'ifico e Tecnol\'ogico, Brazilian research agency, for financial support, grants numbers 303583/2018-5 and 308469/2021-6.
\end{acknowledgements}

%\section*{References}
%  \setcounter{secnumdepth}{0}
%\begin{thebibliography}{99}

%% `Elsevier bibliographyc' style
% \bibliographystyle{elsarticle-num}
% \bibliographystyle{natbib}
% \bibliographystyle{unsrt}
%  \bibliographystyle{plain}
% \bibliography{Liu_HoffdaSilva_Pereira}
%\section*{References}
%\bibliography{main2}
%\bibliography{References}
 %\bibliography{liu}
%%%%%%%%%%%%%%%%%%%%%%%

\end{document}